\documentclass[twocolumn,aps,superscriptaddress]{revtex4-1}
\usepackage{palatino}
\usepackage{lipsum}
\usepackage[utf8]{inputenc}
\usepackage{amsmath}
\usepackage{amssymb}
\usepackage{latexsym}
\usepackage{calc}
\usepackage{color}
\usepackage{graphicx}
\usepackage{hyperref}
\usepackage[none]{hyphenat}

\newcommand{\ben}{\begin{equation}}
\newcommand{\een}{\end{equation}}
\newcommand{\bea}{\begin{eqnarray}}
\newcommand{\eea}{\end{eqnarray}}

\def\sss{\scriptscriptstyle\rm}

\nohyphens 

\def\1s{_{1,\sss S}}
\def\2s{_{2,\sss S}}

\def\bR{{\bf R}}

\def\dulr{{\underline{\underline{\bf r}}}}
\def\dulR{{\underline{\underline{\bf R}}}}
\def\dulP{{\underline{\underline{\bf P}}}}

\def\duld{{\underline{\underline{\bf d}}}}
\def\dulF{{\underline{\underline{\bf F}}}}

\begin{document}
\title{A study of the decoherence correction derived from the exact factorization approach for non-adiabatic dynamics}
\author{Patricia Vindel-Zandbergen}
\email{pv.zandbergen@rutgers.edu}
\affiliation{Department of Physics, Rutgers University, Newark 07102, New Jersey USA}
\author{Lea M. Ibele}
\affiliation{Department of Chemistry, Durham University, South Road, Durham DH1 3LE, UK}
\author{Jong-Kwon Ha}
\affiliation{Department of Chemistry, School of Natural Science, Ulsan National Institute of Science and Technology (UNIST), 50 UNIST-gil, Ulsan 44919, Republic of Korea}
\author{Seung Kyu Min}
\affiliation{Department of Chemistry, School of Natural Science, Ulsan National Institute of Science and Technology (UNIST), 50 UNIST-gil, Ulsan 44919, Republic of Korea}
\author{Basile F. E. Curchod}
\affiliation{Department of Chemistry, Durham University, South Road, Durham DH1 3LE, UK}
\author{Neepa T. Maitra}
\email{neepa.maitra@rutgers.edu}
\affiliation{Department of Physics, Rutgers University, Newark 07102, New Jersey USA}
\date{\today}
\begin{abstract}
We present a detailed study of the decoherence correction to surface-hopping that was recently derived from the exact factorization approach. Ab initio multiple spawning calculations that use the same initial conditions and same electronic structure method are used as a reference for three molecules: ethylene, methaniminium cation, and fulvene, for which non-adiabatic dynamics follows a photo-excitation. A comparison with the Granucci-Persico energy-based decoherence correction, and the augmented fewest-switches surface-hopping scheme shows that the three decoherence-corrected methods operate on individual trajectories in a qualitatively different way, but results averaged over trajectories are similar for these systems. 
\end{abstract}

\maketitle

\section{Introduction}
Trajectory surface-hopping (SH) is one of the most widely-used methods to simulate coupled electron-ion dynamics in molecules~\cite{T90,LAP16,CB18,SJLP16}. While using a classical treatment of the nuclear motion, SH is nevertheless able to capture some quantum features of correlated electron-ion dynamics such as wavepacket splitting, lacking in the Ehrenfest method, another widely-used classical-trajectory based method. SH makes no a priori assumptions regarding relevant degrees of freedom, and, importantly, is relatively straightforward to implement through an interface with electronic structure codes that have the capability to yield excited state energies and gradients.
At the same time, SH has an unsettling aspect, in that there is a disconnect between how the electrons and nuclei evolve, a problem commonly referred to as ``overcoherence": at any given time the nuclei evolve on a single Born-Oppenheimer (BO) potential energy surface, but can instantaneously hop between them according to a stochastic algorithm dependent on the non-adiabatic coupling strengths, while the electronic evolution remains in a coherent superposition of BO states throughout. To overcome this inconsistency, several decoherence corrections have been proposed~\cite{SJLP16,SOL13,SBPR96,PR97, SS11,JFP2012,ZNJT04,GK2008,GP07,GPZ2010}, which, like the SH procedure itself, are somewhat adhoc, even if physically motivated. 

The exact factorization approach~\cite{AMG10,AMG12}, on the other hand, opens the possibility of {\it deriving} a decoherence correction from first-principles since it defines equations for a single nuclear wavefunction and conditional electronic wavefunction that exactly describe the coupled system. 
Ref.~\cite{HLM18} developed a SH scheme with a decoherence correction adopted from the electronic equation derived from a mixed quantum-classical treatment of the exact factorization formalism.
The resulting method, SHXF, has been applied to a number of molecules demonstrating fascinating light-triggered phenomena, like for example the photodynamics of molecular motors or the ring-opening process of cyclopropanone and cyclohexadiene~\cite{FPMC19,FPMK18,FMK19,FMC19}. 

The performance of SHXF has not been compared yet with other decoherence corrections, nor with higher-level non-adiabatic dynamics methods (aside from model systems where exact results are available~\cite{HLM18}). Such comparisons would need some care to be meaningful. In particular, the same initial nuclear geometries and momenta should to be chosen, as well as the same electronic structure method and basis set. Further, it is strongly preferable that the same electronic structure code is used, since, for example, different codes utilize different convergence conditions for self-consistent field calculations that can yield quite different energies and couplings. This can be important especially when molecules evolve far from their equilibrium geometries. 

In this work we study the nature and performance of the SHXF decoherence correction on three molecules for which ab initio multiple spawning (AIMS)~\cite{BQM00,BM98} results are available. AIMS serves as a benchmark: it is based on an expansion of the nuclear wavefunction in terms of coupled trajectory basis functions (multidimensional moving frozen Gaussians), which makes it naturally free from the decoherence issue described in the SH context while yet remaining a trajectory method~\cite{BQM00,CM18,MC18,AC19}. This enables controlled comparisons with surface-hopping.
Two of the molecules, ethylene and fulvene represent two of the recently introduced ``molecular Tully models"~\cite{IC20}, while the third is the methaniminium cation. The latter is chosen because it shares the features of repeated surfaces crossings that  the third molecular Tully model of Ref.~\cite{IC20} has (DMABN), but is easier to explore with different methods due to its smaller size.
For each molecule, a comparison is made with AIMS, with the Granucci-Persico energy-based decoherence correction (SHEDC)~\cite{GP07,GPZ2010}, and with the augmented fewest switches surface-hopping (A-FSSH)~\cite{SOL13,JES16} using precisely the same initial conditions and electronic structure methods. 
We find that the SHXF, SHEDC, and A-FSSH decoherence corrections operate in very different ways on an individual trajectory, but, at least for the systems studied, when averaged over the full set of trajectories, the results for the electronic populations, and nuclear geometry dynamics are very similar. 
We find that in some cases the choice of the velocity-rescaling and/or nuclear time-step have an equal, if not more, important role than the decoherence correction. Finally, implications for further developments of mixed quantum-classical methods are discussed, but first we begin with a brief review of the exact factorization and the SHXF method. 

\section{SHXF}
\label{sec:SHXF}
In the exact factorization approach, the full molecular wavefunction is represented exactly as a single correlated product, $\Psi(\dulr,\dulR,t) = \chi(\dulR,t)\Phi_\dulR(\dulr,t)$ where $\dulr,\dulR$ are all the electronic and nuclear coordinates respectively. The factorization is unique up to a gauge-like transformation, where the nuclear wavefunction $\chi$ is multiplied by an $\dulR$- and $t$-dependent phase while the conditional electronic wavefunction $\Phi$ is multiplied by the inverse phase, provided the partial normalization condition $\int d\dulr \Phi_\dulR(\dulr,t)\vert^2 = 1$ is satisfied. It can be shown that $\chi(\dulR,t)$ reproduces the density and current-density of the nuclear system, and we refer the reader to Refs.~\cite{AMG10,AMG12,AMG13} for more details on the formal properties of the approach, including the relation to the Born-Huang expansion. 

The equations for $\chi(\dulR,t)$ and $\Phi_\dulR(\dulr,t)$ are, not surprisingly, at least as hard to solve as the full molecular TDSE~\cite{GLM19}, however they offer a new starting point for approximations. One such approximation is the coupled-trajectory mixed quantum-classical (CT-MQC) approximation~\cite{MAG15,AMAG16,MATG17,CAT18,GAM18}. This was derived from the exact equations in a particular gauge, and 
taking the classical limit of the nuclear equation; this yields nuclear trajectories that satisfy classical Hamilton-Jacobi equations in a Lagrangian frame. Two further approximations are made to simplify the terms that couple the electronic and nuclear equations, and are well-justified by earlier studies of the exact terms made on model systems~\cite{AMAG16,AASMMG15}.                          
This results in a set of equations that have the form of Ehrenfest plus correction terms that depend on the nuclear quantum momentum, $\nabla \vert \chi \vert/\vert\chi\vert$. Through these terms, the classical nuclear trajectories ``talk" to each other, and result in branching of the electronic coefficients and splitting of the nuclear wavepacket in a consistent way. Decoherence, which in a sense can be viewed as dynamics where the nuclear wavepacket motion is correctly correlated with nuclear-configuration-dependent electronic coefficients, naturally arises. 
CT-MQC has been demonstrated and analyzed on the one-dimensional Tully models~\cite{MAG15,AMAG16,GAM18}, very recently on the photoisomerization of a retinal chromophore model~\cite{MOLA20}, as well as on the process of ring-opening in oxirane~\cite{MATG17,CAT18}, where it was implemented in the CPMD code, interfaced with DFT electronic structure in a plane-wave basis. Regarding computational expense, it is in a sense comparable to surface-hopping: on the one hand it is more expensive because the correction terms involve evolving trajectories and an accumulated force along any BO surface that ever gets populated, but this is compensated by needing far less trajectories to converge as it is not a stochastic method. 
However, while the SH approach is somehow embarrassingly parallel -- each trajectory can be run fully independently -- the formalism of CT-MQC imposes to run the trajectories together, requiring more computational power at the same time and effectively making it significantly slower. The quantum momentum requires input from all trajectories that are being run, i.e. it is not an independent trajectory method. With further computational developments, this impediment may be able to be removed.  

A second mixed quantum-classical approximation, denoted here as SHXF, was developed in Ref.~\cite{HLM18}, in which the electronic equation has the same form as that in CT-MQC but used within a surface-hopping framework with the nuclear trajectories evolving using forces from one BO surface at a time, instantaneously hopping between them according to the fewest-switches hopping algorithm. The correction term appearing in the electronic equation brings about decoherence in a similar way as it did in the CT-MQC algorithm, but is calculated using auxiliary trajectories spawned on non-active surfaces in order to retain an independent trajectory framework. Some details of the algorithm are presented in Sec.~\ref{sec:SHXFeqns}. As mentioned earlier, SHXF has been demonstrated on a range of fascinating processes on complex molecules~\cite{FPMC19,FPMK18,FMK19,FMC19}. 

\subsection{SHXF equations: decoherence and other SH considerations}
\label{sec:SHXFeqns}
In surface-hopping methods an ensemble of classical nuclear trajectories are evolved, $\dulR^{(J)}(t)$, each associated with an electronic wavefunction. The equation that the electronic system satisfies in SHXF is as follows:
\ben
\dot C_n^{(J)} = -\frac{i}{\hbar}\epsilon_n^{(J)}C_n^{(J)} -\sum_k \sum_\nu {\bf d}^{(J)}_{nk,\nu}\cdot{\dot{\bf R}^{(J)}_\nu}C_k^{(J)} + \xi^{(J)}_n
\label{eq:SH}
\een
(with terms all time-dependent), where the last term introduces decoherence, and its form differs between different schemes; for SHXF we have
\ben
\xi^{(J)}_n =  \sum_k \sum_\nu\frac{1}{M_\nu}\left.\frac{\nabla_\nu |\chi|}{\vert\chi\vert}\right\vert_{\dulR^{(J)}(t)}\cdot
\left({\bf f}^{(J)}_{k,\nu} - {\bf f}^{(J)}_{n,\nu}\right) \vert C_k^{(J)} \vert^2 C_n^{(J)}
\label{eq:SHXFe}
\een
Above, $C_n^{(J)}(t)$ denotes the electronic coefficient in the expansion in BO states of the electronic wavefunction associated with the $J$th nuclear trajectory, $\Phi^{J}(\dulr, t) = C_n^{(J)}(t) \Phi_{{\rm BO},n}(\dulr, t)$, while $\epsilon_n^{(J)} = \epsilon_n(\dulR^{(J)}(t))$ is the BO potential energy surface evaluated at the current position of the nuclear trajectory. In the second term of Eq.~\ref{eq:SH},  ${\bf d}^{(J)}_{nk,\nu} = \left.\langle \Phi_{{\rm BO}, n}\vert\nabla_\nu\Phi_{{\rm BO}, k} \rangle\right\vert_{\dulR^{(J)}(t)}$ is the non-adiabatic coupling vector between BO states $n$ and $k$ with $\nu$ labelling the nucleus. The effectiveness of this coupling in causing an electronic transition is dependent on its projection along the nuclear velocity for the $\nu$th nucleus, ${\dot{\bf R}^{(J)}_\nu}$. The third term $\xi^{(J)}(t)$ brings about decoherence, and is given in Eq.~\ref{eq:SHXFe}. This depends on the quantum momentum as well as the accumulated force, i.e. the difference in force along the BO surfaces integrated along the trajectory, ${\bf f}^{(J)}_{k,\nu} = -\int^t\nabla_\nu \epsilon_{{\rm BO},k}^{(J)}(t') dt'$. This term becomes effective when there is some population on more than one state, 
as clear from the dependence on the population factor; for example, if initially the system begins in an excitation to a single electronic excited state, the term is zero, and only gets turned on after the system has evolved near a region of non-adiabatic coupling where some electronic population begins to transfer. The reader is referred to Refs.~\cite{AMAG16,GAM18} for details  on the mechanics of how this term leads to decoherence and wavepacket splitting in model systems.

Turning now to the nuclear equation, we first note that it is the same whether any decoherence correction is applied or not.
For most of the time, the nuclear trajectory follows classical equations of motion along a single BO surface, the ``active" surface, but instantaneously switches surfaces (``hops") according to a prescription that depends in some way on the coupling between the states. The fraction of trajectories in the ensemble that are evolving on the $k$th surface at a given time $t$, $\Pi_k(t) = \sum_J^{N_{\rm traj}}N^{(J)}_{k}(t)/N_{\rm traj}$, defines an electronic population distinct from the population obtained directly from the electronic equation, $\rho_{kk}(t) =  \sum_J^{N_{\rm traj}}\rho^{(J)}_{kk}(t)/N_{\rm traj}$, with  $\rho^{(J)}_{kk}(t)=\vert C^{(J)}_k(t) \vert^2$, and in usual post-calculation analyses, it is $\Pi_k(t)$ that is ultimately recorded as the electronic population, while $\rho_{kk}(t)$ is disregarded. 
In the fewest-switches scheme~\cite{T90}, an expression for the hopping probability algorithm was developed by considering the requirement of ``internal consistency": that is, the average over the ensemble of many trajectories, $\Pi_k(t)$ should be equal to the average $\rho_{kk}(t)$, while minimizing the number of hops. However, since SH is run with independent trajectories, these averages are not available, and instead the expression is applied in a stochastic sense to the individual trajectories, which breaks the internal consistency~\cite{GP07}. 
The resulting stochastic algorithm depends on the hopping probability between the active state $a$ and another state $\zeta_{ak}$:
\ben
\zeta_{ak}^{(J)} = \max\left\{0,-\frac{2\Re (\rho_{ak}^{(J)*} {\bf d}^{(J)}_{ka,\nu}\cdot{\dot\bR^{(J)}_\nu}) }{\rho_{aa}^{(J)}} dt, \right\}
\een
where $\rho_{ak}^{(J)} = C_a^{(J)*} C_k^{(J)}$.
Then, a hop from the active state $a$ to the state $n$ is made if $\sum_{k=1}^{n-1} \zeta_{ak}^{(J)} < r \le \sum_{k=1}^{n} \zeta_{ak}^{(J)}$ where $r$ is a random number uniformly distributed in $[0, 1]$.

 The violation of internal consistency in pure SH (i.e. Eq.~\ref{eq:SH} with $\xi = 0$) is fundamentally due to combining fully coherent electronic coefficient evolution with nuclear dynamics that in contrast evolves on a single surface at any given time, jumping surfaces stochastically. There is thus a disconnect. The nuclear trajectory in the electronic equation is the same for the coefficient associated with any surface even though the forces as defined from the gradient of the  different surfaces are different. Further, frustrated hops (see Sec.~\ref{sec:veladj}) exacerbate the problem. Adding the decoherence correction $\xi(t)$ acts to push the electronic coefficients to the active state, dampening them on the non-active surfaces. As mentioned before, the SHXF correction can be derived from the exact factorization equations. 
 
 We briefly discuss some key aspects of how the SHXF correction is computed; full details can be found in Ref.~\cite{HLM18}. 
 To retain an independent trajectory description, auxiliary trajectories are used to evaluate the quantum momentum appearing in the decoherence term in the SHXF equation~\cite{HLM18}. For each independent trajectory,  an auxiliary trajectory is generated on the non-active surfaces when the population of that surface becomes non-zero (or above a small threshold). The auxiliary trajectory is launched with a velocity such that the difference in potential energy from the active surface is isotropically distributed in the coordinates, and this velocity then steps forward the position of the auxiliary trajectory. 
In this way, the calculation of gradients of auxiliary surfaces is avoided, aiding in computational efficiency. In a similar spirit, the accumulated force along a surface is calculated from directly computing the change in momentum over a time-step. 
 The quantum momentum is obtained by considering a Gaussian of isotropic width $\sigma$ centered at each auxiliary trajectory; from which follows that the quantum momentum is given by the distance of the average of the auxiliary trajectory positions, weighted by the populations, to the actual trajectory's position. 
 
There is clearly a significant numerical cost reduction in using auxiliary trajectories to compute the quantum momentum instead of actually coupling the different surface hopping trajectories. A price to pay for this is the introduction of the parameter $\sigma$. We avoid empiricism by fixing it to be the width of the ground-state nuclear wavepacket at the initial equilibrium geometry.


 \subsubsection{Other decoherence schemes}
We will compare the effect of the SHXF $\xi(t)$ on the dynamics to two widely-used decoherence corrections, SHEDC and A-FSSH, which we now briefly discuss.

The SHEDC decoherence correction has quite a different form to SHXF, acting directly on non-active states to damp the amplitude on them at a rate that depends on the energy gap 
$\epsilon_{{\rm BO}, n}(\dulR^{(J)}(t)) - \epsilon_{{\rm BO}, a}(\dulR^{(J)}(t))$
between the surfaces, and the kinetic energy $T$ of the nuclei~\cite{GP07,GPZ2010,ZNJT04,ZJT05}. 
It is imposed as an exponential decay of amplitudes on the non-active state, which, if written in the form of Eq.~\ref{eq:SHXFe}, would correspond to an effective 
\ben
\xi^{(J), {\rm SHEDC}}_{n\neq a}(t) =-\frac{|\epsilon_{{\rm BO}, n}^{(J)} - \epsilon_{{\rm BO}, a}^{(J)}|}{\hbar}\left(1 + \frac{\alpha}{T}\right)^{-1} C^{(J)}_{n\neq a}
\een
while for the active state $a$, the coefficient is adjusted so that the sum of all coefficients is 1. 
The parameter $\alpha$ is a constant, and could be adjusted but mostly is fixed as $0.1$ H~\cite{ZNJT04}. It should be noted that although the original papers proposed to apply this decay to the populations, in some versions of widely-used codes, such as the one we use here, the correction is applied to the coefficients. However, numerical comparisons between the two approaches for a subset of molecules do not reveal significant practical differences in the results~\cite{IC20}.

In another contrasting approach, A-FSSH defines a decoherence rate based on considering how fast trajectories evolving on different surfaces move away from each other~\cite{JES16}; this was motivated by a comparison with the quantum-classical Liouville equation~\cite{SOL13}. Each trajectory carries with it auxiliary trajectories evolving on different surfaces, which are propagated classically, similar to SHXF. In A-FSSH, however, the electronic coefficient is collapsed to a state in a stochastic manner, as determined by a decoherence rate computed from
\ben
	\frac{1}{\tau^{\rm A-FSSH}_{na}} =   \frac{\delta {\dulF}_{n} \cdot \delta \dulR_{n} }{2\hbar}   - \frac{2\vert \duld_{an} \cdot\dot\dulR (\epsilon_{{\rm BO},a} - \epsilon_{{\rm BO},n})\delta \dulR_{n}\cdot\dot\dulR\vert}{\hbar \vert \dot\dulR\vert ^2}
\een
where, 
$\delta\dulR_n=\dulR_n-\dulR_a$ is the position of the trajectory on auxiliary surface $n$ relative to the position of the trajectory on the active surface $a$, $\delta{\dulF}_n=-\nabla_\nu(\epsilon_n(\dulR_n) - \epsilon_a(\dulR_a))$ is the difference in BO forces on surface $n$ and $a$, and, everywhere in the equation the dot product means e.g. $\duld_{an}\cdot \dot\dulR = \sum_\nu {\bf d}^{(J)}_{an,\nu}\cdot\dot\bR_\nu^{(J)}$ and $\vert\dot\dulR\vert^2=\sum_{\nu}\vert \dot\bR_\nu\vert^2$. 
If we were to write this as an effective decoherence term in Eq.~\ref{eq:SHXFe}, we would have $\xi_{n \neq a}^{(J), {\rm A-FSSH}} = -C^{(J)}_{n}/\tau^{\rm A-FSSH}_{n\neq a} $. However, the rate is instead used in a stochastic procedure: if $dt_c/{\tau^{\rm A-FSSH}_{n \neq a}}$ is larger than a random number then the amplitude $C_n$ is collapsed to zero on state $n$ while that on the active state is increased so that the sum of the coefficients remains 1. A separate reset rate is used to then reset $\delta \dulR_n$ to 0.

The three decoherence corrections, exact-factorization derived SHXF, the energy-based SHEDC, and the stochastic coefficient collapse of A-FSSH could not appear more different! Indeed, we will find in Sec.~\ref{sec:Results} that in practise, the way that the three decoherence corrections above act on the trajectories is very different. Still, after averaging over the surface-hopping trajectories, the populations and geometries (not shown here) are similar. 

We next turn to some other issues that any surface-hopping algorithm, decoherence-corrected or not, must confront. 
 
\subsubsection{Convergence questions}
The stochastic hopping process means that several trajectories for each initial condition should be run, and convergence to a given standard error has to be monitored carefully. It requires typically tens to hundreds of trajectories per degree of freedom~\cite{T90,plasser2019strong}. Further, there is the question of the time-step required for convergence: the hopping probability at a given time-step clearly decreases linearly as the nuclear time-step $dt$ decreases, however the system is interrogated whether it wants to hop correspondingly more often, so that it is believed that these two effects compensate. However for very localized avoided crossings or conical intersections, the hopping can be missed unless $dt$ is taken too small to be practical; how many electronic time-steps are used within this $dt$ is also an important factor, including how the interpolation for the electronic propagation is done within $dt$. Using a wavefunction overlap-based approach with local diabatization to obtain the couplings can improve the numerical stability~\cite{HT94,ML14,PGPBPL12,WP14}.
Ref.~\cite{PS20} very recently showed that the stochastic algorithm tends to overestimate the hopping rate when the hopping probability is large, and instead a modified scheme based on a cumulative hopping probability rather than the instantaneous one was proposed that significantly reduces the sensitivity to the time-step, as well as requiring less trajectories for convergence.

\subsubsection{Velocity Adjustment}
\label{sec:veladj}
The SH algorithm in itself lacks a firm first-principles derivation (although see Ref.~\cite{LZ18} for recent progress), and as a consequence, there are aspects of the nuclear dynamics which need to be adjusted in some way. One important aspect is the velocity adjustment after a hop. It is asserted that each trajectory should satisfy energy conservation, where the gain or loss in the potential energy is compensated by a loss or gain in the kinetic energy, but there is no unique way to achieve this~\cite{CGB17}. Two common ways are isotropic rescaling and rescaling along the non-adiabatic coupling vectors (NACV) between the two states ${\bf d}_{an}$. 
We note here that in other trajectory-based schemes where the trajectories are coupled rather than independent, such as in AIMS or CT-MQC, energy conservation of an individual trajectory would not be required. In AIMS, the nuclear velocities of a newly-spawned trajectory basis function is scaled per default along the NACV. AIMS was shown to be insensitive to the rescaling process -- isotropic rescaling produces similar results as the NACV one~\cite{IC20}. 

 In isotropic rescaling, every velocity after the hop is scaled uniformly such that the total energy is conserved: with $\nu$ labelling the atom, $\dot\bR_\nu \to \kappa \dot\bR_\nu$ where $\kappa = \sqrt{1 - (\epsilon_{{\rm BO},n} - \epsilon_{{\rm BO},k})/T}$ and the trajectory hops from surface $k$ to surface $n$. 
Rescaling along the NACV is believed to be theoretically more justified from semiclassical arguments~\cite{Herman84,Pechukas69,CX95}. In this case, $\dot\bR_\nu \to \dot\bR_\nu + \gamma {\bf d}_{\nu,kn}/M_\nu$ where $\gamma$ is determined by the quadratic equation resulting from equating the sum of the nuclear kinetic and potential energy on surface $k$ to that on surface $n$.
Recent work~\cite{SZVPKST20} has shown that both approaches of velocity adjustment lead to a violation of the conservation of angular momentum, which could be overcome by using a projection operator to remove the translational and rotational components of the NACV and rescale the velocity along this projected NACV. 

If the potential energy gain after the hop exceeds the kinetic energy, then the hop is rejected. In this case, some works argue that the nuclear momentum should then be reversed but other works argue that it should be kept as is~\cite{CGB17,HT94,JT03,plasser2019strong}. 
There are generally more rejected (a.k.a. frustrated) hops when rescaling along the NACV is done, since only the kinetic energy along the NACV is available, and this can result in a poorer internal consistency; moreover, the NACV is not always accessible from the electronic structure code being used. On the other hand, a disadvantage of isotropic scaling is that it is size-extensive: even if the dynamics involves just a few atoms of a large molecule or cluster, the rescaled velocity unphysically depends on the entire kinetic energy even of atoms that are not involved in the process. 
These factors suggest a third rescaling procedure: scale via NACV, and when the hop is forbidden, then apply isotropic scaling. 
We refer to this as ``NACV+iso" in the following sections. 
(In fact, the rescaling option denoted as ``NACV" in the Newton-X code does NACV+iso, while the corresponding option in SHARC, which we use in this work, does NACV.)


\section{Computational Details}
\label{sec:Results}
The SH and SHXF, 
calculations are performed with the code PyUNIxMD (UNIversal eXcited state Molecular Dynamics)~\cite{unixmd}. The current capabilities include BO, Ehrenfest, SH and SHXF dynamics, interfaced with a range of electronic structure programs. Since the main objective of the present work is to compare the effect of the decoherence correction derived from exact factorization with SHEDC, A-FSSH, and against the high-level AIMS method which we consider in this work as a reference, we keep other aspects of the calculations the same as much as possible. In particular, for the electronic structure we use CASSCF implemented in MOLPRO~\cite{molpro} for our calculations on ethylene (SA(3)-CASSCF(2/2)), methaniminium cation (SA(2)-CASSCF(6/5)) , and fulvene (SA(2)-CASSCF(6/6)) with the 6-31G$^*$ basis set.
The SHEDC and A-FSSH computations are done with the code SHARC 2.0 (Surface Hopping including Arbitrary Couplings)~\cite{SHARC,MMG18,sharc_prog}.

The initial conditions for the nuclear coordinates and velocities are taken exactly the same as in the AIMS calculations~\cite{IC20}, which is Wigner-sampled from uncoupled harmonic oscillators of frequencies determined from the optimized ground-state geometry of the molecule. For ethylene and the methaniminium cation, both geometries and momenta were sampled from this distribution, while for fulvene just the geometries were Wigner-sampled and initial momenta were set to zero. Every trajectory was averaged using different random seeds to enable the convergence of the FSSH stochastic process; the total number of trajectories for each molecule is detailed below.

The nuclear time-step is taken as $dt = 0.5$ fs unless otherwise stated. We have checked that decreasing the time-step does not alter the results except for the case of fulvene; the convergence is generally  better for the decoherence-corrected schemes than uncorrected.
As will be discussed, the dynamics in fulvene is somewhat sensitive to the choice of time-step. The large slope of the crossing region means that a large number of trajectories encounter the sharp and localized non-adiabatic coupling.

 
For SH and SHXF, 
the explicit NACV were used in the equation of motion, while for A-FSSH and SHEDC they were obtained from wavefunction overlaps by default in SHARC~\cite{PRMOMG16}. We checked there is little difference in the results when using these two approaches, except for the fulvene molecule where the convergence with respect to time-step is better using the wavefunction overlap scheme, as mentioned earlier. An isotropic velocity adjustment was performed after a surface hop unless otherwise stated.

The population trace for AIMS were taken from Ref.~\cite{IC20} for ethylene and fulvene. For the methaniminium cation, AIMS dynamics were performed with the MOLPRO/FMS90 interface~\cite{levine2008implementation}, using an adaptive time-step of 20 a.u. (5 a.u. in regions of non-adiabatic coupling) and a SA(2)-CASSCF(6/5)/6-31G$^*$ level of theory for the electronic structure (mirroring the electronic structure used for the mixed quantum/classical methods). The AIMS parent TBFs were started from the same set of initial conditions as the other non-adiabatic methods.

\section{Results}
Our main objective is to compare the effects of the decoherence correction arising from the exact factorization to the widely-used SHEDC and A-FSSH.

\subsection{Ethylene}
As discussed in Ref.~\cite{IC20}, dynamics after photoexcitation to the $S_1$ state represents a molecular Tully-1 system, since it proceeds through a single non-adiabatic event through a conical intersection. This represents a cis-trans-like isomerization of the molecule through a twisted and pyrimidalized geometry~\cite{BQM00,BRL05}. The importance of having consistent initial conditions and electronic structure methods in comparing different dynamics methods for this molecule were emphasized in Ref.~\cite{IC20}, and here we use the same 66 initial conditions, geometries and momenta, used there, from the Wigner-sampled ground-state geometry. We  ran 10 trajectories for each initial condition but note that results were essentially converged even with 5 trajectories per initial condition. 
The width of the Gaussian, $\sigma$, is obtained from the initial distribution of the nuclear trajectories of the CC double bond, and it is set to $0.05$ a.u.

\begin{figure}
	\includegraphics[width=\linewidth]{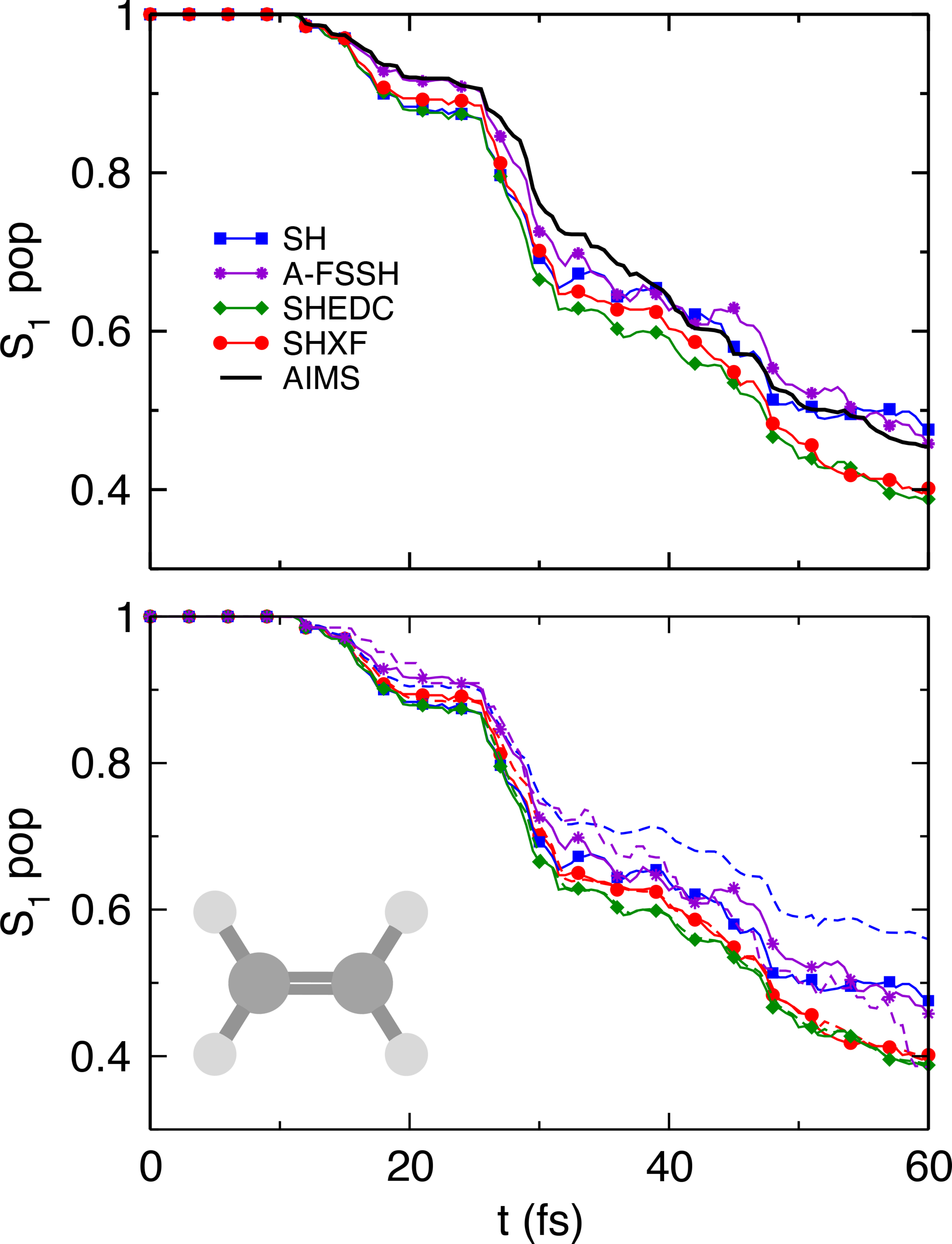}
\caption{Population dynamics in ethylene: SHXF compared with SH, SHEDC and A-FSSH, all with isotropic velocity adjustment, 
	along with the reference AIMS results (from Ref.~\cite{IC20}). The top panel shows the fraction of trajectories $\Pi_{S_1}(t)$ in the $S_1$ state. The lower panel demonstrates the internal consistency of the surface-hopping methods, with the solid lines showing  $\Pi_{S_1}(t)$ again, compared with dashed lines showing the $S_1$ electronic populations $\rho_{S_1,S_1}(t)$.}
\label{fig:eth1}
\end{figure}

In Fig.~\ref{fig:eth1}, we plot the $S_1$ populations as determined by both the fraction of trajectories and the electronic populations, computed from the SH, SHXF, SHEDC and A-FSSH simulations. 
For this system, the fraction of trajectories predicted by uncorrected SH is very close to the reference AIMS, but we see that there is a notable internal consistency error, as expected. Averaged over trajectories, the SHXF decoherence correction from exact factorization and SHEDC yield very similar results, increasing the population transfer compared to the uncorrected SH, and correcting the internal consistency of the uncorrected SH (the electronic populations are practically on top of the fraction of trajectories in both cases). The appear to agree less well with AIMS but do not deviate too far and would lie within the standard error of AIMS~\cite{IC20}. 
A-FSSH is closer to AIMS, but it shows worse internal consistency than SHEDC and SHXF. 

\begin{figure}
	\includegraphics[width=\linewidth]{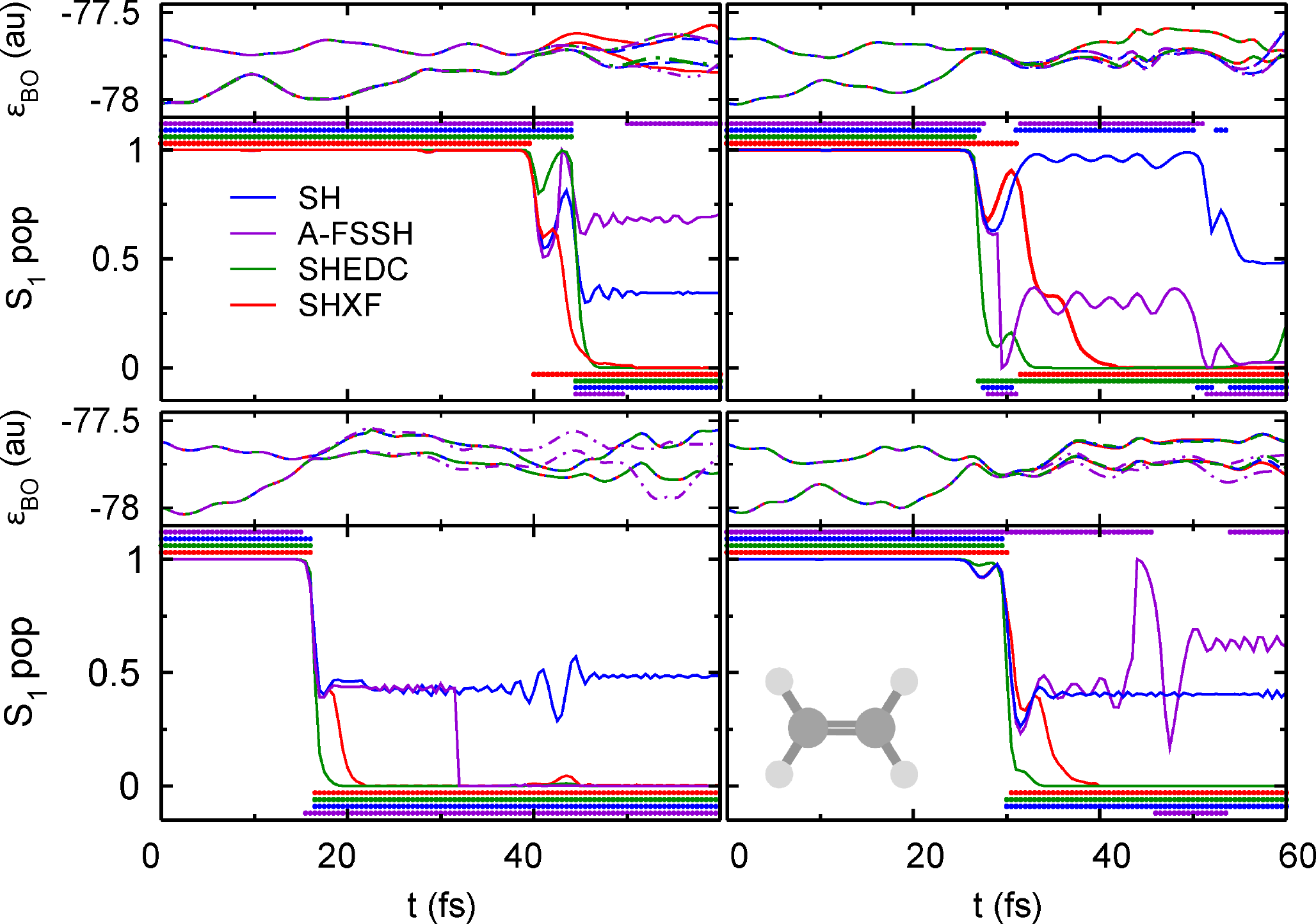}
\caption{Comparing population dynamics in ethylene for 4 trajectories with the same initial conditions, SH, SHXF, SHEDC and A-FSSH, with isotropic velocity adjustment. Continuous lines show the populations $\rho_{S_1,S_1}(t)$ while the correspondingly colored symbols indicate the active state. Top panels show the electronic energies during SHXF dynamics. 
The Appendix gives an AIMS analog for this.}
\label{fig:eth2}
\end{figure}

The close agreement of SHXF, SHEDC and A-FSSH is not obvious, given the different structure of the corrections discussed earlier. Indeed, on an individual trajectory level, their behavior is quite different. In Fig.~\ref{fig:eth2} we show the populations and active state for four randomly chosen trajectories in the SH, SHXF, SHEDC and A-FSSH simulations. The SHEDC correction damps down the populations after a hop in a mostly (but not entirely) monotonic way, while the SHXF tends to be typically non-monotonic, showing more oscillations and generally takes longer to decohere. The stochastic nature of the A-FSSH decoherence correction is clearly evident in the plots, and suggest, for this molecule, a longer decoherence time than the other methods. The Appendix provides an analog to this figure for the AIMS calculations, including a discussion highlighting essential differences between surface-hopping methods and the AIMS approach. 

The different behavior on an individual trajectory level is reflected in an average over all trajectories of the decoherence indicator~\cite{MAG15,AMAG16,MATG17}, defined as $\rho_{10}(t) = \sum_J^{N_{\rm traj}} \vert C_{S_1}^{(J)} C_{S_0}^{(J)} \vert^2/N_{\rm traj}$. The SHXF dynamics grows to a larger coherence, and takes a longer time to decohere than SHEDC, but the overall structure is similar. The coherence peak around 17 fs reflects a small number of trajectories that reach a conical intersection earlier than those associated with the second peak around 30 fs. On the other hand, as clear from the sample trajectories, A-FSSH remains coherent longer. Although in the present case, this difference does not affect the overall population dynamics very much, nor the nuclear geometries (not shown), it opens the  question of whether the different behavior results in other systems.

\begin{figure}
\includegraphics[width=\linewidth]{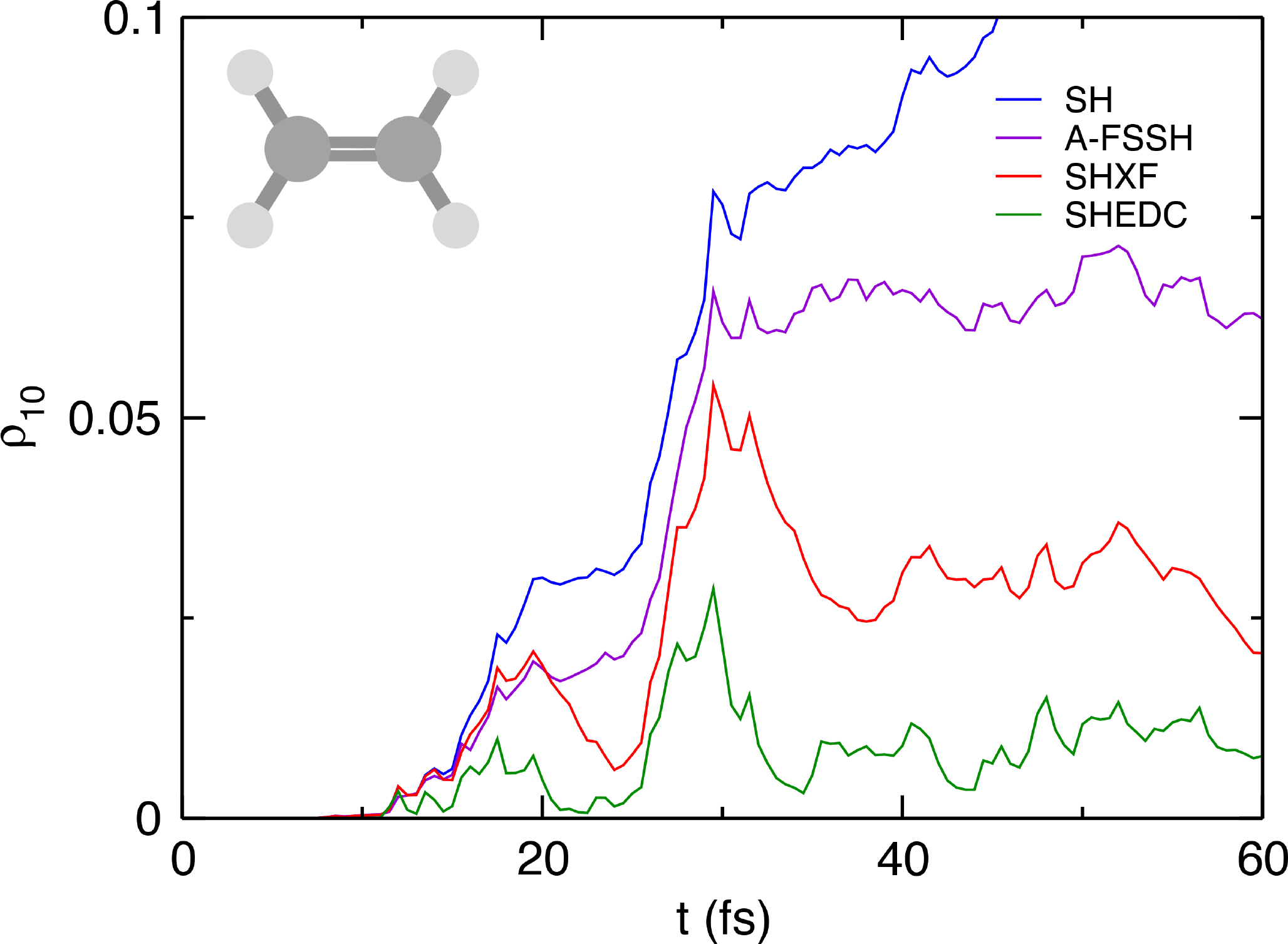}
\caption{Decoherence indicator in ethylene: SH, SHXF, SHEDC and A-FSSH.}
\label{fig:eth3}
\end{figure}

\begin{figure}
\includegraphics[width=\linewidth]{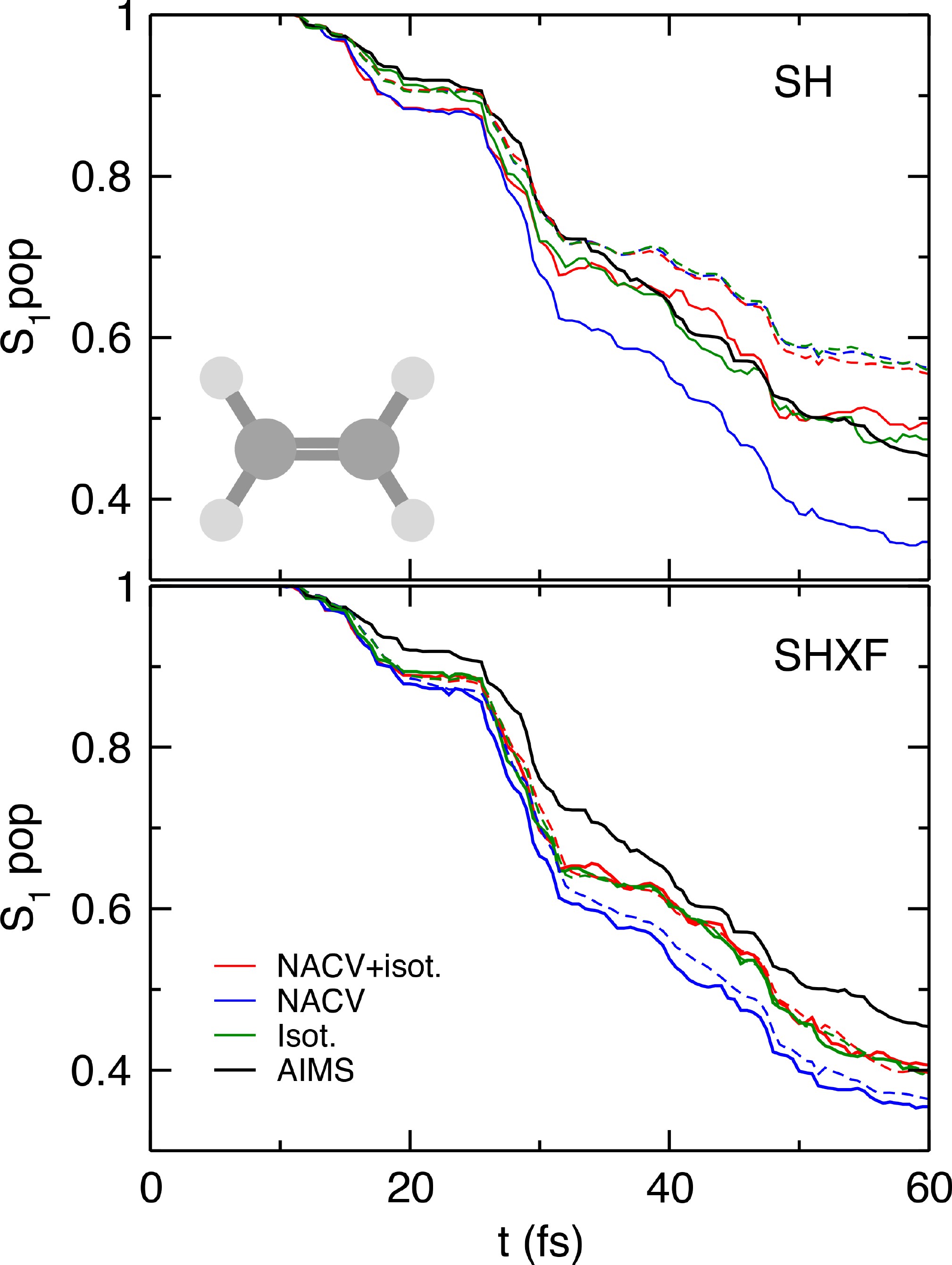}
\caption{Comparison of different velocity adjustments in ethylene. Top panel: uncorrected SH, $\Pi_{S_1}(t)$ and $\rho_{S_1,S_1}(t)$, with velocity adjustments of isotropic, NACV, and NACV-iso; Lower panel: the same with SHXF. 
  AIMS is shown as reference.}
\label{fig:eth4}
\end{figure}

Finally, the importance of the choice of velocity adjustment is evident in Fig.~\ref{fig:eth4}, where the top panel shows the results of uncorrected SH with three different ways of velocity adjustment and the lower panel shows the SHXF case. The spread in the results shows that, in this case, the choice of velocity adjustment has just about as much effect on the dynamics as the decoherence correction. In particular, while the internal consistency is very well corrected by the decoherence correction when using isotropic scaling, errors remain when scaling along NACVs is performed, consistent with the expectation from the earlier discussion on velocity adjustment. When isotropic scaling is used as a ``back-up" to scaling along the NACV in the NACV+iso approach, the error in the internal consistency is again small when the decoherence correction is applied; the results are close to the isotropic scaling case for this molecule.

\subsection{Methaniminium cation}
Despite its apparent similarity to ethylene (isoelectronic and planar but here with a CN double bond), the dynamics of the methaniminium cation after photoexcitation to $S_1$ is quite different: following initiation of the photoisomerization after the excitation the methaniminium cation typically meets another region of non-adiabatic coupling in a different region of configuration space, displaying recrossings with $S_1$ before decaying to $S_0$~\cite{BGPRVEL07}.   
The molecule tends to show torsional motion, and the initial transfer of population to $S_0$ occurs  once the system rotates around the CN bond from 0 to 90 degrees. (This contrasts with the photodynamics obtained by exciting the molecule to $S_2$ electronic state, where bond elongation couples with rotation~\cite{BGPRVEL07}). Here we use 96 initial conditions, each repeated 4 times. 
Preliminary trajectory runs indicate that a time-step of $0.25$ fs leads to converged results with respect to time-step.
The parameter $\sigma$ is set to $0.056$ a.u., which is the uniform variance obtained from the initial distribution of the CN bond of the nuclear trajectories.

\begin{figure}
\includegraphics[width=\linewidth]{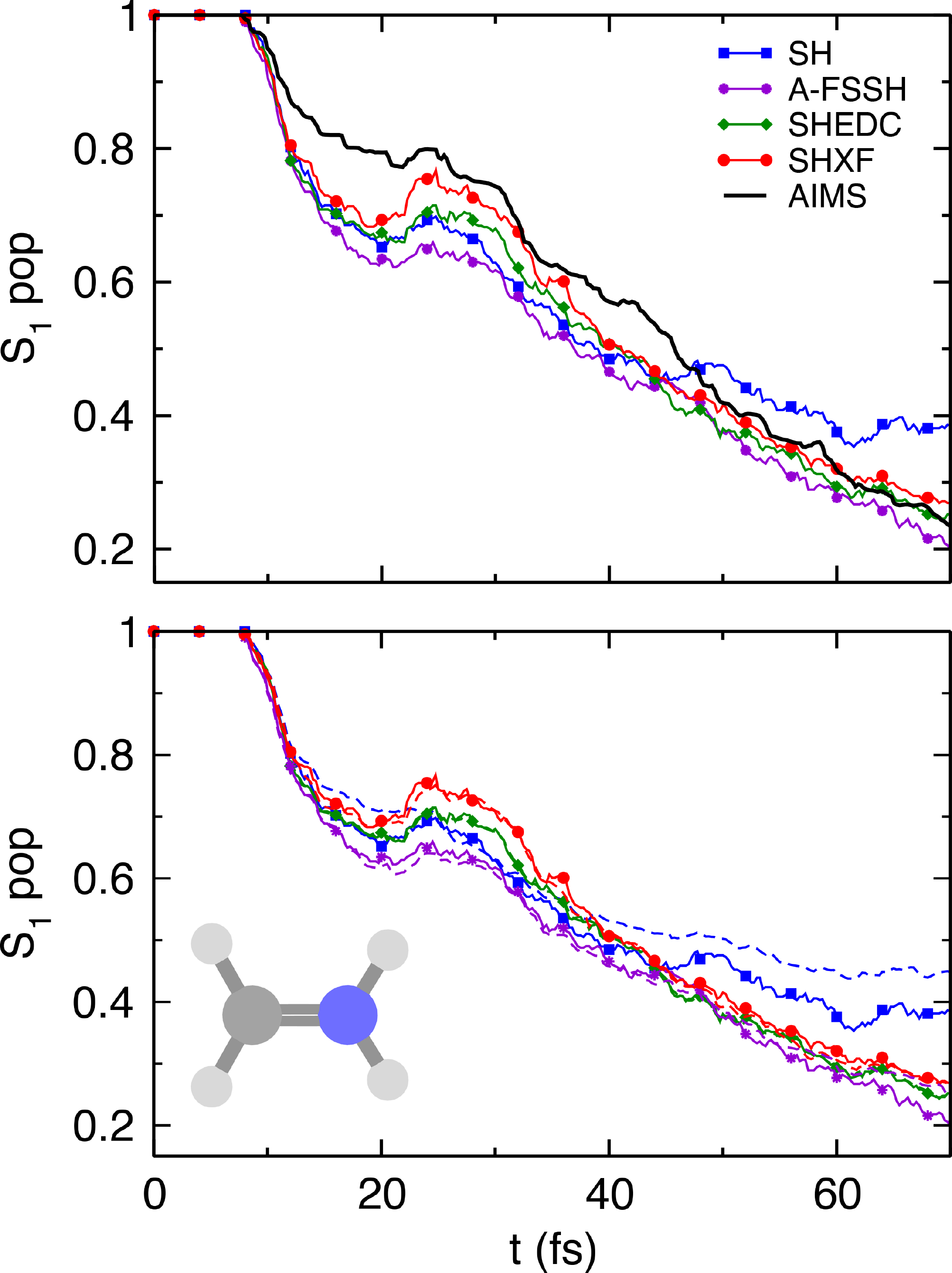}
	\caption{Population dynamics in methaniminium cation: 
	SHXF compared with SH, SHEDC and A-FSSH, all with isotropic velocity adjustment, 
	along with the reference AIMS result (from Ref.~\cite{IC20}). The top panel shows the fraction of trajectories $\Pi_{S_1}(t)$ in the $S_1$ state. The lower panel demonstrates the internal consistency of the surface-hopping methods, with the solid lines showing  $\Pi_{S_1}(t)$ again, compared with dashed lines showing the $S_1$ electronic populations $\rho_{S_1,S_1}(t)$.}
\label{fig:meth1}
\end{figure}

Figure~\ref{fig:meth1} shows the population dynamics in SHXF as compared with SH, SHEDC and A-FSSH, all using isotropic velocity adjustment, with the reference AIMS shown. After some fast transfer around $10$ fs, where the molecule initiates a direct photoisomerization to $S_0$, the populations then plateau with  recrossings back to $S_1$ before then steadily transferring to $S_0$, as mentioned earlier.
 
The poor internal consistency of the uncorrected SH is evident after the first transfer, and especially at later times. The overcoherence of uncorrected SH impacts the populations at later times, yielding less transfer to $S_0$ than AIMS and  the decoherence-corrected SH methods. 
The decoherence-corrected methods all correct this, particularly well for SHEDC. They give reasonable agreement with AIMS, and capture the $S_0 \to S_1$ population transfer back around $25$ fs, with SHXF the most enthusiastic. 
The initial population transfer to $S_0$ is however too fast especially for A-FSSH.

\begin{figure}
\includegraphics[width=\linewidth]{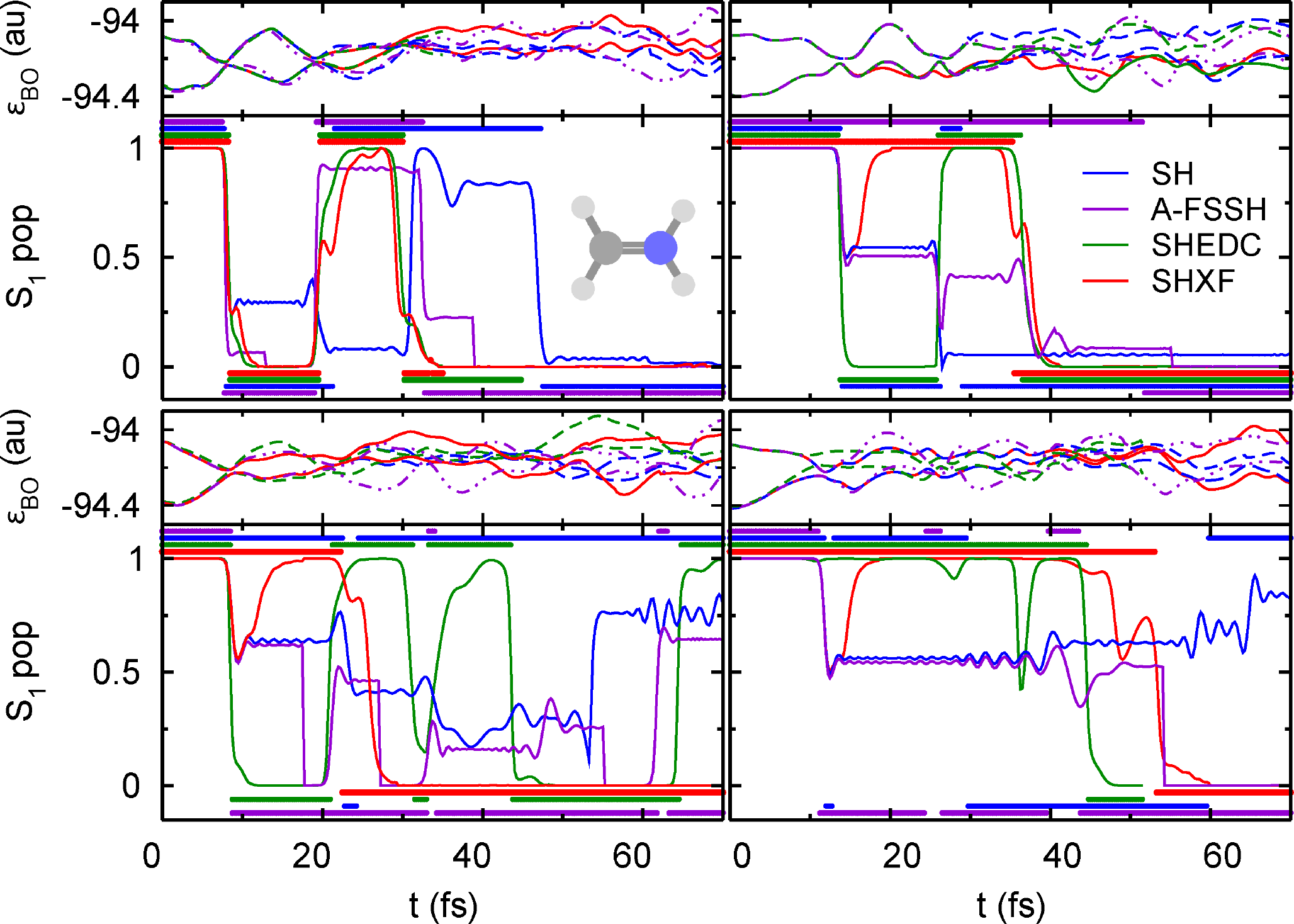}
\caption{Comparing population dynamics in methaniminium cation for 4 trajectories with the same initial conditions, SH, SHXF, SHEDC and A-FSSH, with isotropic velocity adjustment. Continuous lines show the population $\rho_{S_1,S_1}(t)$ while the correspondingly colored symbols indicate the active state. Top panels show the electronic energies during SHXF dynamics. 
}\label{fig:meth2}
\end{figure}

\begin{figure}
\includegraphics[width=\linewidth]{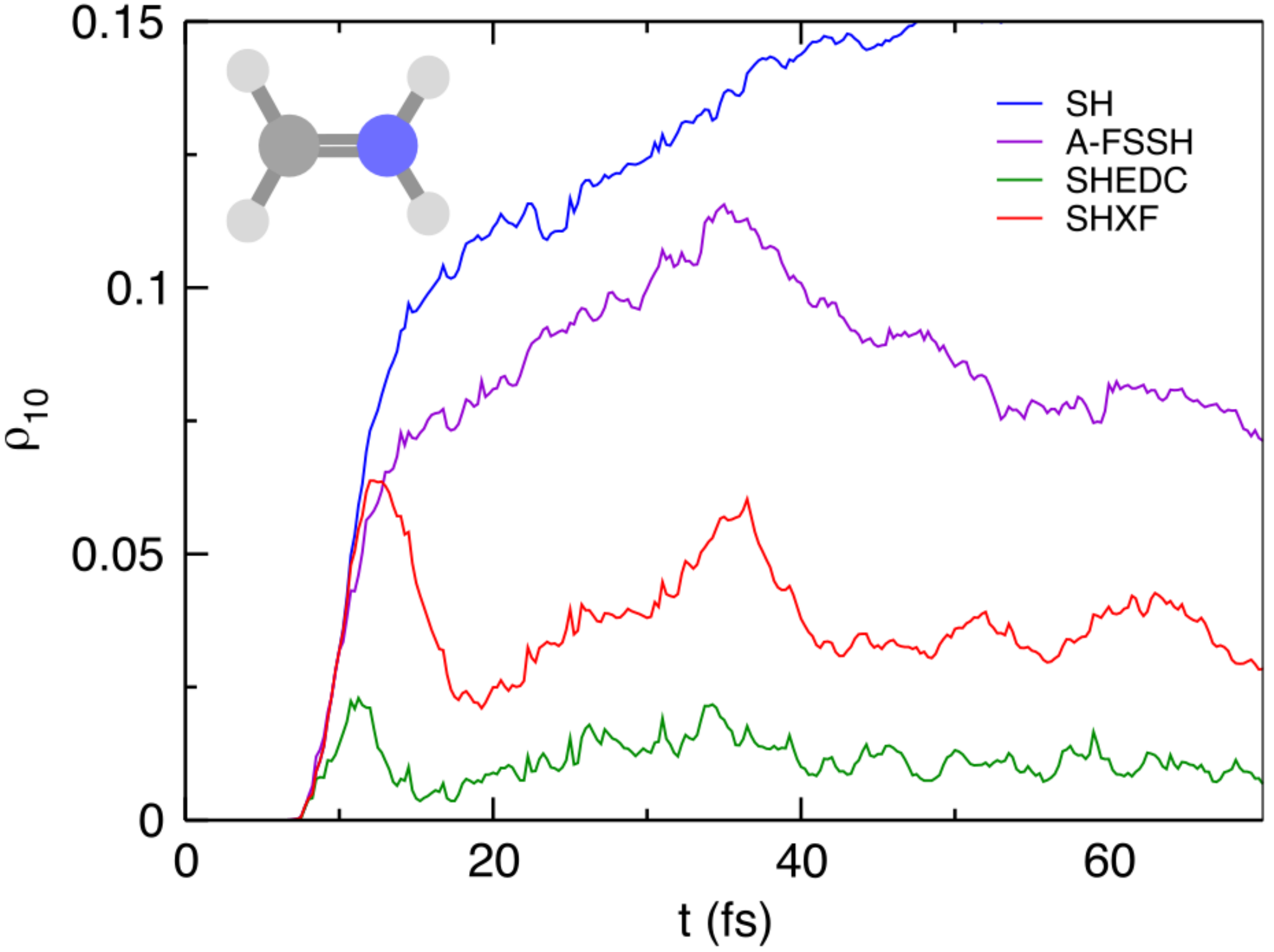}
\caption{Decoherence indicator in methaniminium cation: SH, SHXF, SHEDC, and A-FSSH}
\label{fig:meth3}
\end{figure}

Again on an individual trajectory level, the decoherence corrections act in different ways on the electronic populations, as evident from the sampling of trajectories shown in Fig.~\ref{fig:meth2}, and this is again reflected in the trajectory-averaged quantity, the decoherence indicator, shown in Fig.~\ref{fig:meth3}. Again SHXF shows a similar coherence structure to SHEDC but reaches larger values, while A-FSSH is somewhat different and takes longer to decohere. 
Fig.~\ref{fig:meth2} also highlights further the recrossings between $S_0$ and $S_1$ states, leading to a more severe deviation of SH from internal consistency (Fig.~\ref{fig:meth1}) than for ethylene.

\subsection{Fulvene}
Fulvene represents a challenging case: 
After photoexcitation to $S_1$ state, there are two possible pathways for an ultrafast internal conversion to the ground state~\cite{MLWBR10,IC20,ILMC21}. One involves a peaked conical intersection reached by a
a twist of the C=CH$_2$ bond, while the other involves a strongly sloped conical intersection reached by the stretch of the C=CH$_2$ bond~\cite{IC20}. The latter results in a transfer to $S_0$ and subsequent reflection back towards the same non-adiabatic region and population transfer back to $S_1$ state. This second pathway resembles the Tully model III and, as in Ref.~\cite{IC20}, we choose the initial conditions to favor this. The $\sigma$ parameter is chosen as $0.065$ a.u., which corresponds to the variance of initial distribution of CC double bonds of the nuclear trajectories.

\begin{figure}
\includegraphics[width=\linewidth]{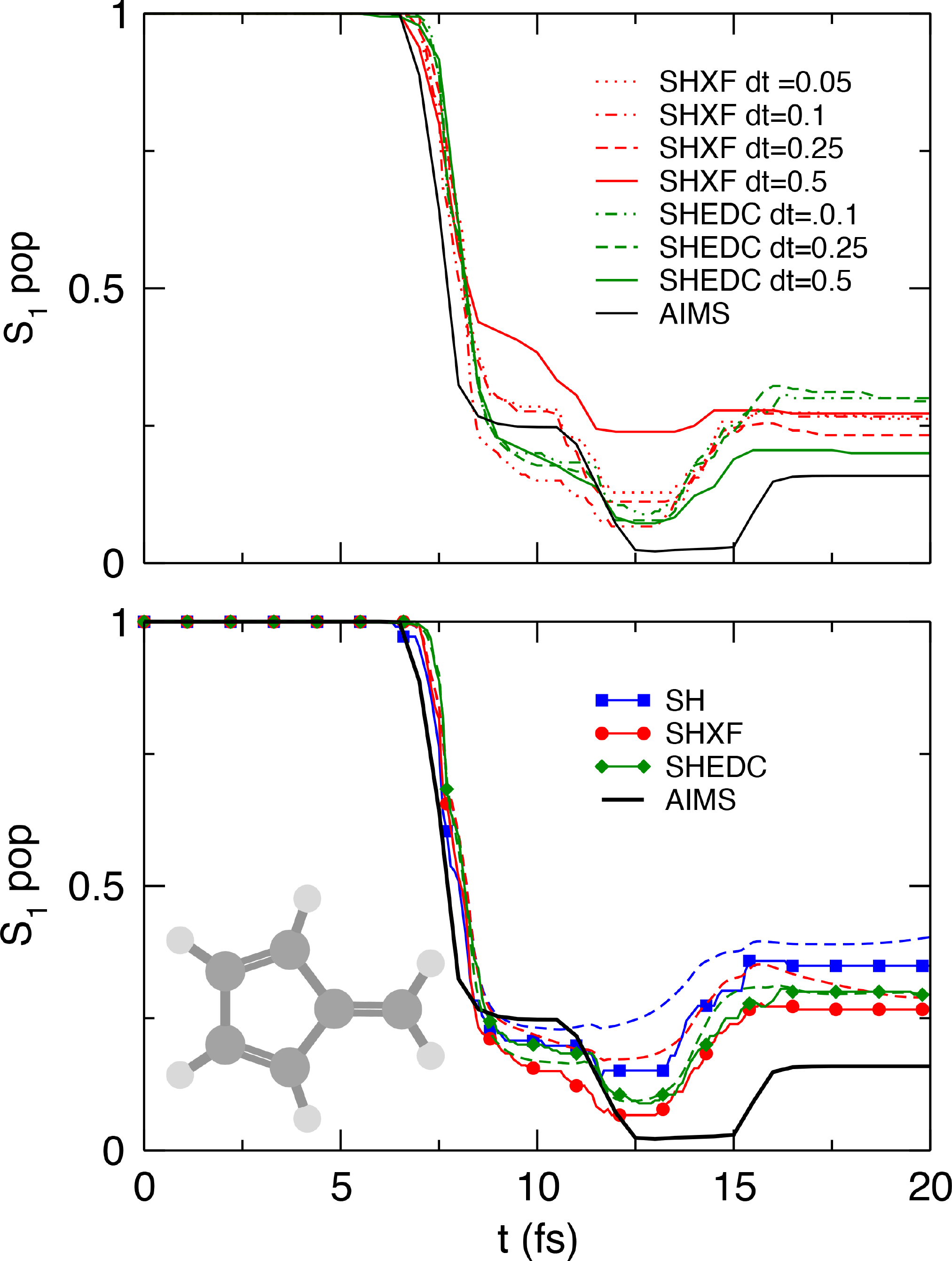}
\caption{Fulvene populations. Top panel shows the convergence of $\Pi_{S1}$ with respect to the nuclear time-step $dt= 0.05$, $0.1$, $0.25$, $0.5$. 
Lower panel: choosing $dt = 0.1$, we plot the fraction of trajectories $\Pi_{S1}$ along with $\rho_{S_1,S_1}$ (dashed), for SHXF, SHEDC, and SH against the AIMS reference.}
\label{fig:fulvdt}
\end{figure}

The sharply sloped conical intersection gives a large dependence on the time-step $dt$, since the interaction region can be missed. 
We see that as $dt$ decreases from $0.5$ fs to $0.25$ fs to $0.1$ fs, SHXF predicts more population during the initial event (Fig.~\ref{fig:fulvdt}, top panel) but that the $dt = 0.05$ fs result is closer to the $dt = 0.25$ fs result than to the $dt = 0.1$ fs result; the results are thus not fully converged with respect to the time-step.  
To some degree, this dependence can be mitigated by using wavefunction-overlaps to compute the coupling terms, with a local diabatization scheme. The SHEDC calculations in SHARC utilize this scheme, and we see in the top figure that although SHEDC predictions with $dt = 0.5$ fs (green dash-dot line) plateau to a different level after $15$ fs (and is closer to the AIMS result) than that predicted with the $dt = 0.1$ fs and $0.25$ fs calculations, the results do appear converged with $dt = 0.25$. This example highlights the need to check for convergence with respect to the time-step in these cases. As mentioned earlier, the recent method of Ref.~\cite{PS20} is promising in this regard.  We note that AIMS uses an adaptive time-step so does not have such sensitivity. 

In the lower panel we see that both decoherence-corrected schemes increase the population transfer compared to pure SH, with good internal consistency. Both SHEDC and SHXF agree quite well with each other, despite their different operation mechanisms. 

Finally, it was observed in Ref.~\cite{IC20} that the dynamics heavily depends on the choice of velocity-adjustment. Isotropic scaling gives results notably worse than scaling along the NACV for this molecule, which might be explained due to the larger size of the molecule, since the problem with unphysical redistribution of the kinetic energy in the isotropic method becomes more important. The results shown in Fig.~\ref{fig:fulvdt} used scaling along the NACV. 

\section{Conclusions}
\label{sec:Concs}
Overall, the results show that SHXF provides a useful improvement over uncorrected surface-hopping in comparison with the reference AIMS, and gives a similar behavior for  observables as SHEDC and A-FSSH. 
The three decoherence corrections suggest strikingly different mechanisms on an individual trajectory level. This was clear in both the form of the corrections, as well as their demonstrated behavior on the molecular systems. 
For the systems studied, the different decoherence mechanisms nevertheless yielded similar population dynamics once averaged. This seems unlikely to be true generically, given their different modes of operation. Whether one can somehow predict when the differences will lead to significantly different observables, and why they were so similar here,  is a question for future research. 

Several adhoc aspects of the SH approach itself, arising from the fact that SH itself is not an algorithm derived consistently from first-principles makes it difficult to give a definitive and unambiguous performance of the corrections themselves, and in some cases issues such as velocity scaling  procedures, for which different procedures have been argued to be best,  give larger differences than the decoherence corrections themselves. Thus, in parallel to further exploring SHXF and its capabilities -- especially for large systems given its computational efficiency, further developments of CT-MQC and alternative practical mixed quantum classical methods from the exact factorization is an avenue for future work.

\begin{figure}
\includegraphics[width=\linewidth]{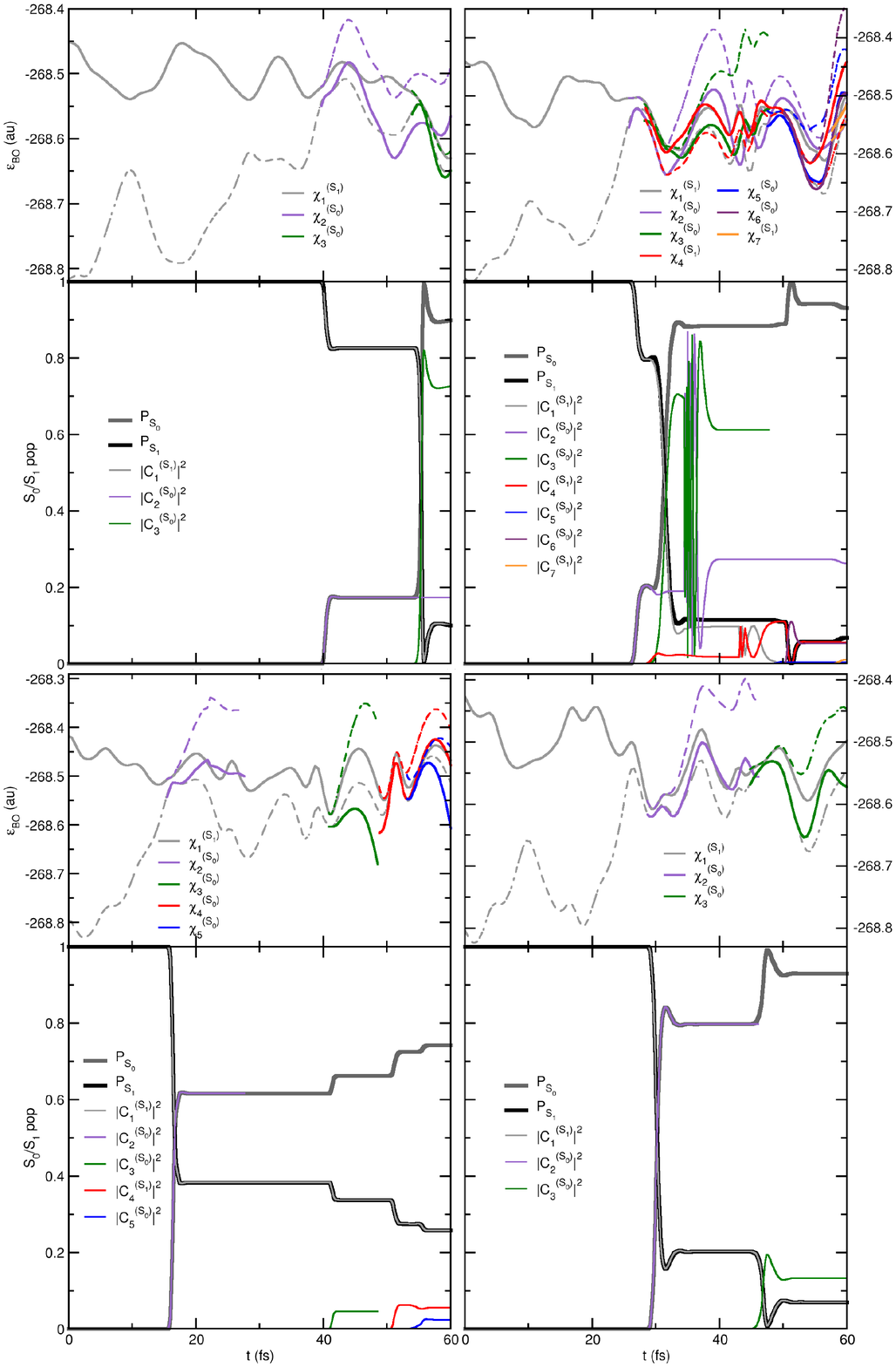}
	\caption{Comparing the AIMS population dynamics in ethylene for four different initial conditions (same as those presented in Fig.~\ref{fig:eth2}). The top panels show the electronic states of all TBFs, where the bold line shows the electronic energy of the BO state in which the TBF evolves while the dashed lines indicate the electronic energy of the other electronic state. The bottom plots show the evolution of the population of the states ($P_{S_0}(t)$ and $P_{S_1}(t)$ as defined in Eq.~\eqref{eqpop}, shown with thick, black and grey lines), as well as the evolution of the squared modulus of each TBF amplitude ($|C_J^{(k)}(t)|^2$).}
\label{fig:aimstbfs}
\end{figure}


\appendix
\section{Analysis of AIMS runs for ethylene}

We present here an AIMS analog of Fig.~\ref{fig:eth2} for ethylene. In AIMS, the nuclear wavefunction for each BO state is described by a linear combination of frozen Gaussians, the so-called trajectory basis functions (TBFs), 
\begin{equation}
\begin{aligned}
\chi^{(k)}&(\dulR, t)= \\
&\sum_k^{N_{T}^{(J)}(t)}C_J^{(k)}(t)\tilde{\chi}_J^{(k)}\left(\dulR;\overline{\dulR}_J^{(k)}(t),\overline{\dulP}_J^{(k)}(t), \underline{\underline{\boldsymbol{\alpha}}}, \overline{\gamma}_J^{(k)}(t) \right) \, ,
\end{aligned}
\end{equation}

where $\tilde{\chi}_J^{(k)}\left(\dulR;\overline{\dulR}_J^{(k)}(t),\overline{\dulP}_J^{(k)}(t), \underline{\underline{\boldsymbol{\alpha}}}, \overline{\gamma}_J^{(k)}(t) \right)$ are multidimensional Gaussians, each associated with a time-dependent complex coefficient $C_J^{(k)}(t)$, where $J$ labels a specific TBF, evolving in electronic state $k$. The phase-space center of each multidimensional Gaussian functions is given by $\overline{\dulR}_J^{(k)}(t)$ and momentum $\overline{\dulP}_J^{(k)}(t)$. The matrix $\underline{\underline{\boldsymbol{\alpha}}}$ contains the widths (same for all TBFs and independent of the electronic state) and $\overline{\gamma}_J^{(k)}(t)$ is a phase. The TBFs evolve along classical trajectories and the spawning algorithm will increase the size of the TBFs basis when nonadiabatic regions are encountered (see Ref.~\cite{CM18} for additional details on AIMS).

An AIMS calculation starts with one parent TBF, assigned to a selected electronic state and with a given set of initial conditions for the nuclear positions and momenta. One can follow the electronic energy of the driving state along the dynamics of the parent TBF. This is given by a plain grey line, noted $\chi_1^{(S_1)}$) in Fig.~\ref{fig:aimstbfs} ($J=1$ as it is the first TBF and $k=S_1$). The dashed line with same color represents the electronic energy for $S_0$, along the TBF evolving on $S_1$. When the TBF reaches a region of strong nonadiabaticity, a new TBF is spawned onto the coupled state, here $S_0$, and evolves with nuclear forces given by the electronic ground state (noted $\chi_2^{(S_0)}$ in Fig.~\ref{fig:aimstbfs}). In other words, the second TBF will have its own dynamics in $S_0$, and deviate from that of the parent TBF -- compare the dashed grey line ($S_0$ energies on the support of $\chi_1^{(S_1)}$) with the plain blue line ($S_0$ energies on the support of $\chi_2^{(S_0)}$). We stress here that the parent TBF $\chi_1^{(S_1)}$ still exists and carries on its dynamics on $S_1$, as seen from the plain grey curve. The spawning process will be repeated every time a TBF reaches a region of strong nonadiabaticity, increasing the number of TBFs ($N_{T}^k(t)$) to describe the nuclear wavefunction in $S_0$ and $S_1$.

The previous paragraph described how the TBFs evolve on the different PESs, in other words, how the moving adaptive grid spreads over time. We now need to discuss how the TDSE is solved on the support of these TBFs. This is achieved by solving the TDSE in the basis of the TBFs, leading to coupled equations of motion for the complex coefficients $C_J^{(k)}(t)$. At the beginning of the dynamics, the parent TBF $\chi_1^{(S_1)}$ is assigned a complex coefficient $C_1^{(S_1)}(t_0)=(1.0\,\,0.0)$. Following a spawn, the newly created TBF $\chi_2^{(S_0)}$ carries initially a complex coefficient $C_2^{(S_0)}(t_\text{entry})=(0.0\,\,0.0)$. ($t_\text{entry}$ is the time when the parent TBF originally triggered the spawning mode, when the nonadiabatic couplings crossed a certain predefined threshold.) The coefficients are coupled via the TDSE and can exchange nuclear amplitude, as observed in Fig.~\ref{fig:aimstbfs}. We note that the population of a given electronic state is not equal to the summation of the population on each TBF evolving on this state, due to the non-orthogonality of the multidimensional Gaussians. Instead, one can get the actual AIMS population in state $S_0$, $P_{S_0}$ by calculating the expectation value of the projector $\hat{\mathcal{P}}_{S_1} = | \Phi_{S_1} \rangle\langle \Phi_{S_1} |$ using the AIMS molecular wavefunction:

\begin{widetext}
\begin{equation}
\begin{aligned}
	P_{S_1}(t)  &  = \sum_{kn}^\infty \Big[ \sum_J^{N_{T}^k(t)}\sum_{I}^{N_{T}^n(t)}
\left(C_{J}^{(k)}(t)\right)^\ast C_{I}^{(n)}(t)  \langle   \Phi_k \tilde{\chi}_{J}^{(k)} |  \hat{\mathcal{P}}_{S_1}  | \tilde{\chi}_{I}^{(n)} \Phi_n \rangle_{\dulr, \dulR}
\Big]  \\
& =  \sum_{JI}^{N_{T}^{S_1}(t)}\left(C_{J}^{({S_1})}(t)\right)^\ast C_{I}^{({S_1})}(t)  \langle \tilde{\chi}_{J}^{({S_1})}  | \tilde{\chi}_{I}^{({S_1})}  \rangle_{\dulR}  =   \sum_{JI}^{N_{T}^{S_1}(t)} \left(C_{J}^{({S_1})}(t)\right)^\ast C_{I}^{({S_1})}(t)  S^{{S_1}{S_1}}_{J,I} \, . 
\label{eqpop}
\end{aligned}
\end{equation}
\end{widetext}

In Eq.~\eqref{eqpop}, $S^{{S_1}{S_1}}_{J,I}$ is an overlap matrix element between TBFs $J$ and $I$. The AIMS populations are given by thick lines in Fig.~\ref{fig:aimstbfs}.


Fig.~\ref{fig:aimstbfs} also highlights the conceptual difference between AIMS and SH. As every newly created TBF evolves independently, decoherence is naturally accounted for. In addition, AIMS assures at the individual trajectory level a much smoother population transfer, as it does not rely on instantaneous hops but merely on Gaussians that will interact and have the possibility to transfer population between each other continuously. Indeed, all initial conditions show a stepwise deactivation process in AIMS, where multiple spawns are required. Interestingly, in one of the cases (top right plots of Fig.~\ref{fig:aimstbfs}) a small repopulation of the $S_1$ state can be observed, mediated by back spawns to that state. In contrast, such effects are not reproduced in the corresponding SH trajectories, as these are just minor population transfers that only a sufficiently large swarm of SH trajectories would capture.

\acknowledgments
This work was primarily supported by the Computational Chemical Center: Chemistry in Solution and
at Interfaces funded by the U.S. Department of Energy, Office of Science, Basic Energy Sciences under Award No.
DE-SC0019394 (PVZ) as part of the
Computational Chemical Sciences Program. This grant also applies for the calculations carried out on Temple Univesity's HPC resources. Partial support from the Department
of Energy, Office of Basic Energy Sciences, Division of Chemical
Sciences, Geosciences and Biosciences under Award No. DESC0020044 (NTM) is also acknowledged. 
This project has received funding from the European Research Council (ERC) under the European Union's Horizon 2020 research and innovation programme (Grant agreement No. 803718, project SINDAM). LMI acknowledges the EPSRC for an EPSRC Doctoral Studentship (EP/R513039/1).
PV and NTM thank Spiridoula Matsika for useful conversations.

\bibliography{./pnlb6_4.bib}

\end{document}